\documentstyle[12pt]{article}
\begin{document}

\title{A globally accurate theory for a class of binary mixture models}
\author{Adriana G. Dickman$^{a}$ and
G. Stell$^{b}$\\
$^a$ {\small Departamento de F\'\i sica e Qu\'\i mica,
Pontif\'\i cia Universidade Cat\'olica de Minas Gerais,}\\
{\small Av. Dom Jos\'e Gaspar, 500,
Cora\c c\~ao Eucar\'\i stico, 30535-610, Belo Horizonte - MG, Brasil}\\
$^b$ {\small Department of Chemistry, State University of New York at Stony Brook,
Stony Brook,}\\{\small New York 11794-3400, U.S.A.}}
\date{\today}
\maketitle
\begin{abstract}

Using the self-consistent Ornstein-Zernike approximation (SCOZA) results for
the 3D Ising model, we obtain phase diagrams for binary mixtures
described by decorated models.
We obtain the plait point, binodals, and closed-loop coexistence
curves for the models proposed by Widom, Clark, Neece, and Wheeler.
The results are in good agreement with series expansions and experiments.

\vspace{2.0truecm}

\end{abstract}
\vspace{1.0truecm}

\noindent $^a${\small electronic address: adickman@pucminas.br} \\
\indent{\small phone number: 55-31-3319-4176}\\
$^b${\small electronic address: gstell@mail.chem.sunysb.edu} \\
\indent{\small phone number: +1-631-632-7898}\\
\indent{\small fax number: +1-631-632-7960}\\

\newpage

\centerline{\bf 1. Introduction}

\vskip 12pt

Critical phenomena of mixtures are considerably more complicated to investigate
than such phenomena in pure fluids. Purification of components,
precise determination of composition, gravitational effects and long
equilibration times make experiments on mixtures much more difficult to
realize.
Similar difficulties are found in theoretical analyses, where the addition
of one component greatly complicates the
generalization of the lattice-gas model to mixtures, for example the
derivation and
analysis of series expansions \cite{wheeler}.

The use of decorated lattice-gas models to describe fluid mixtures
represents an
important gain in precision and information about these models. This is
a result of the exact mapping between these models and the
single-species lattice-gas
or Ising model. The exact results in two dimensions, and exhaustive series
analysis in three, allow a detailed description of the phase diagrams,
coexistence surfaces and critical exponents of the much more complicated
models for mixtures.

In this paper we obtain information related to critical phenomena
of binary mixtures for various decorated lattice models using Ising-model
results
obtained via the SCOZA \cite{scoza}.
This approximation provides very precise results for the lattice-gas Ising
model in three dimensions as shown elsewhere \cite{scoza2}.

The next sections define the models we work with and show how to
derive the grand partition function for these models. Then we present a
brief discussion about SCOZA. Finally, we present the phase diagrams and
coexistence curves we obtain for the binary mixture models.

\vskip 12pt

\centerline{\bf 2. Decorated model}

\vskip 12pt

Consider a binary mixture of $N_1$ molecules of type 1 and $N_2$ molecules
of type 2 contained in a volume $V$. A decorated lattice model for a mixture
is constructed as follows \cite{wheeler,widom67}. We divide the volume
into C primary cells, which are centered on sites of a regular lattice, and
secondary cells centered on the bond midpoints of this lattice.
If the primary cells have volume $v_0$, then the secondary cells
occupy a volume $2v_0/q$, where $q$ is the lattice coordination number.
Each primary cell overlaps $q$ secondary cells and each secondary cell
overlaps two primary cells, as shown in figure~1. A primary cell
is occupied if a type-1 molecule lies within its boundaries; secondary cells
are occupied by type-2 molecules.
The cells are not meant to restrict the molecules to lattice sites.
They merely provide a discrete coordinate system for measuring
the positions of particles of type 1 and 2 and thus defining intermolecular
separations.

\centerline{[Insert figure~1 about here]}

All interactions are pairwise. If two molecules of type 1
are in the same primary cell, the interaction energy is $+\infty$; this
energy is $-\epsilon$ if they occupy adjacent primary cells, and zero
otherwise. The interaction energy for
a pair of type-2 molecules is $+\infty$ if they occupy the same secondary
cell and zero otherwise. If a primary cell containing a type-1 molecule
overlaps a secondary cell occupied by a type-2 molecule, the interaction
energy is $\phi$. Particles of different species have no interaction if
their cells do not overlap.

The condition that there is no interaction between type-2 molecules in
different secondary cells is essential to the existence of an exact
mapping between the decorated model and the spin-$1\over 2$ Ising model.
As a consequence, the pure component 2 cannot undergo a phase transition
\cite{wheeler}.

So far we have described the decorated model in general terms. The specific
models
for binary mixtures by Widom, Clark, Neece, and Wheeler are obtained by
adjusting the parameters for the interaction energy.

Widom \cite{widom67} defined the simplest model by setting a hard-core
repulsion between molecules of types 1 and 2, the only interaction beyond
exclusion of
multiple occupancy. The model undergoes a first-order phase transition even
though the intermolecular forces are all assumed to be infinitely strong
repulsions.

Clark \cite{clark68} generalized Widom's model by allowing a finite repulsion
between unlike nearest-neighbour molecules. All other interactions between
like molecules are zero and multiple occupancy of cells is not allowed.
This model has a coexistence dome with a maximum critical solution temperature.

Neece \cite{neece67} extended Widom's model including the effects of
interactions between molecules on the primary lattice. This model exhibits
a variety of behaviours with changing temperature, especially if the molecules
on the primary lattice interact with an attractive potential.
It provides a good qualitative description of certain
types of `gas-gas equilibrium' such as that found in the He-Xe system.

Wheeler \cite{wheeler72} studied the binary case with an attractive interaction
between molecules on the primary lattice, and a finite interaction (attractive
or repulsive) between type-1 and -2 molecules. This version proved useful in
understanding critical phenomena in very dilute solutions.

Another model studied by Wheeler \cite{wheeler75} considers a highly
directional, short-ranged interaction favoring special orientations of the
molecules at low temperatures. Unlike the other models, a primary or secondary
lattice site can be occupied either by a type-1 or a type-2 molecule.
Interactions between particles of the same
species (1-1 or 2-2) are zero regardless of their orientations. The 1-2
interaction depends upon the orientation of the particle in the secondary
cell. If the 1-2 bond involves the special contact point of this particle
then the energy is negative (attractive) otherwise it is positive (repulsive).
Each molecule (1 or 2) has one special contact point and $(\omega-1)$ other
identical contact points regardless of whether it is on a primary or secondary
cell. We consider the case with $\omega = 6$.
Thus, if a molecule of either type is in a secondary cell which has one
1-filled and one 2-filled primary neighbour, then it has one orientation with
energy $U_2 <0$ and five orientations with energy $U_1>0$. If both their
primary neighbours are of the opposite type, then there are two orientations
in which it makes one attractive and one repulsive contact with resulting
energy $U_2+U_1$, and four orientations in which it makes two repulsive
contacts
with resulting energy $2U_1$. The 1-2 interaction energy is independent of the
orientation of the particle in the primary cell. This is required to map
the model
exactly onto the spin-1/2 Ising model. This model presents closed-loop
coexistence curves with both upper and lower critical solution temperatures,
analogous to the behaviour found in the nicotine + water and m-toluidine +
glycerol systems.

Table~1 shows a summary of the models presented above.

\centerline{[Insert table~1 about here]}

\vskip 12pt
\centerline{\bf 3. Derivation of the Grand Partition Function}
\vskip 12pt

In this section we derive the grand partition function for the
decorated models and show how to map these models into the lattice gas
Ising model.
We follow the derivation given by Wheeler (for details see \cite{wheeler}).
The grand partition function can be written as
\begin{eqnarray}
{\cal Z}=\sum_{{\cal C}_1}\sum_{{\cal
C}_2}z_1^{N_1}x^{N_{11}}z_2^{N_2}e^{-{E^{(2)}\over kT}}
\label{grand1}
\end{eqnarray}
where the sums are over all assignments of filled or empty primary
and secondary cells. $E^{(2)}$ is the sum
of all primary-secondary interactions, $N_{11}$ is the number of filled-filled
nearest neighbour pairs of primary cells and $x=\exp(-\epsilon/kT)$.
$z_1$ and $z_2$ are dimensionless activities defined as,
\begin{eqnarray}
z_1 &=& v_0\left({2\pi m_1 kT\over h^2}\right)^{d\over 2}e^{\mu_1\over kT},\\
z_2 &=& \left({qv_0\over 2}\right) \left({2\pi m_2 kT\over h^2}\right)^{d\over 2} e^{\mu_2\over kT},
\label{z12}
\end{eqnarray}
where $d$ is the dimensionality.

Let $\eta_{ij} = e^{-\psi_{ij}/kT}, i,j=0,1$, where $\psi_{00}, \psi_{01},
\psi_{11}$ are the energies for a molecule of type 2 that lies between two
empty, one
empty and one filled, or two filled primary cells, respectively. The idea
is to sum over
filled and empty states for each secondary cell, once a specific
configuration of
the primary cells is chosen. Thus if a secondary cell lies between empty-empty
primary cell pairs, it contributes to the grand partition function a factor
$1+z_2\eta_{00}$, if it lies between empty-filled primary cell pairs, it
contributes a factor $1+ z_2\eta_{01}$, and if it lies between filled-filled
primary cell pairs, it contributes a factor $1+z_2\eta_{11}$. Then,
\begin{eqnarray}
{\cal Z}=\sum_{{\cal C}_1}z_1^{N1}x^{N_{11}}(1+z_2\eta_{00})^{N_{00}}
(1+z_2\eta_{01})^{N_{01}}(1+z_2\eta_{11})^{N_{11}}
\label{grand2}
\end{eqnarray}
where $N_{00}$ and $N_{01}$ are the number of empty-empty and empty-filled
nearest neighbour primary cells. Using the lattice identities,
$qN_1 = 2N_{11} + N_{01}, qN_0 = 2N_{00} + N_{01}, N_1 \equiv N$ and $N_0 =
C - N$
we get
\begin{eqnarray}
{\cal Z}=(1+z_2\eta_{00})^{qC\over 2}\sum_{{\cal C}_1}z^{N} \exp\left({\epsilon
N_{11}\over kT}\right)=
(1+z_2\eta_{00})^{qC\over 2}{\cal Z}_R.
\label{grand3}
\end{eqnarray}
Comparing this with the partition function for the lattice gas,
\begin{eqnarray}
{\cal Z}_R = \sum_{\cal C}~(z_R)^N \exp({\beta_R~\epsilon_R}N_{11}),
\nonumber
\end{eqnarray}
we obtain the relations that map the decorated model exactly to the lattice gas
Ising model:
\begin{eqnarray}
{\cal Z} &=& (1+z_2\eta_{00})^{qC\over 2}{\cal Z}_R \nonumber\\
z_R &=& z_1(1+z_2\eta_{00})^{-q}(1+z_2\eta_{01})^q \nonumber \\
\exp\left({\epsilon_R\over kT_R}\right) &=&
x(1+z_2\eta_{00})(1+z_2\eta_{11})(1+z_2\eta_{01})^{-2}.
\label{mapping}
\end{eqnarray}
The subscript $R$ denotes that the quantity refers to the reference system.
This is an exact correspondence between the decorated model and the spin-$1/2$
Ising model. All thermodynamic properties can be obtained from the grand
canonical potential $\pi(z_1,z_2,x) = \lim_{c\rightarrow\infty}{1/C}\ln{\cal Z}$
by differentiation.

The thermodynamic pressure $p(\mu,T)$ is given by
\begin{eqnarray}
\beta p = \lim_{V\rightarrow\infty} {1\over V} \ln{{\cal Z}(T,\mu,V)}=
{q\over 2}\ln{(1+z_2\eta_{00})}+\beta_R~p_R
\label{pressure}
\end{eqnarray}
for the decorated model. The number densities are given by:
\begin{eqnarray}
\rho_1 &=& \left({\partial\pi\over \partial\ln z_1}\right)_{z_2,x} = \rho_R\nonumber\\
\rho_2 &=& \left({\partial\pi\over \partial\ln z_2}\right)_{z_1,x} = {q\over 2} A +
q(C-A)\rho_R + (A+B-2C){\Phi_R\over \epsilon_R}
\label{densities}
\end{eqnarray}
where $\Phi_R$ is the energy density in the reference system and,
\begin{eqnarray}
A={z_2\eta_{00}\over (1+z_2\eta_{00})}~~
B={z_2\eta_{11}\over (1+z_2\eta_{11})}~~
C={z_2\eta_{01}\over (1+z_2\eta_{01})}.
\label{abc}
\end{eqnarray}
Other quantities of interest are the internal energy, given by
$U=(\partial\pi/\partial\beta)_{z_1,z_2}$, and the specific heat at
constant volume, given by $C_{v,x}=(\partial U/\partial T)_{\rho_1,\rho_2}$.

\vskip 12pt
\centerline{\bf 4. Self-Consistent Ornstein-Zernike Approximation(SCOZA)}
\vskip 12pt

The SCOZA closure of the Ornstein-Zernike equation
yields critical parameters of remarkable accuracy for three-dimensional
lattice gases with nearest-neighbour interactions \cite{scoza,scoza2}.

In the SCOZA,
the direct correlation function $c(r)$ is determined in a way that results
in a unique
free energy function.
The starting point is the Ornstein-Zernike relation
\begin{eqnarray}
h_i = c_i + \rho\sum_{Lattice} c_{i-j}h_j
\end{eqnarray}
where $h$ is the total correlation function. Thermodynamic consistency is
embodied in the exact relation that must be satisfied by the derivation
of the free energy
\begin{eqnarray}
\rho {\partial^2(\rho u)\over \partial\rho^2} = {\partial^2(\beta p)\over
\partial\beta\partial\rho}
\end{eqnarray}
where $u$ is the internal energy per particle, $p$ is the pressure and $\rho$
the density.
$c_0\equiv c(0)$ is fixed by the core condition on the total correlation
function, $h_0 = -1$.
The single approximation of the method is the truncation of the direct
correlation
function $c(r)$ at the nearest-neighbour separation: $c(r) = 0$ for $r>1$.
From the relations between $c_0, c_1$ and $h_1$ we find a partial
differential equation for $c_1(\beta,\rho)$ that can be solved numerically.
For details see references~\cite{scoza,scoza2}.

\vskip 12pt
\centerline{\bf 5. Results}
\vskip 12pt

In this section we specify the mapping equations for each model of interest and
we also show the corresponding phase diagrams. We study binary mixtures in
three dimensions. Thus the
reference one-component lattice gas has a range of thermodynamic states
corresponding to liquid-vapor coexistence terminating in a critical
point. At the boundary of the two-phase region in the ($\Phi/\Omega, N/\Omega$)
plane of the reference system the activity $z_R$, the temperature $T_R$,
and the pressure-temperature ratio $p_R/T_R$ are all singular, having
discontinuous derivatives there. Therefore, we see from equations
(~\ref{mapping}) and (\ref{pressure}),
that the functions $z_1,z_2,p/T$ of the two-component models are singular at
the corresponding point in the ($\rho_1,\rho_2$) plane.

As a reminder, in our notation $\epsilon$ is the interaction energy between
two type 1 molecules when sitting at neighbouring primary cells, and $\phi$
denotes
the interaction energy  between a type-1 and a type-2 molecule when they
occupy overlapping primary and secondary cells.

\vskip 12pt
\centerline{5.1. Reference system}
\vskip 12pt

Our reference system is the lattice gas transcription of the Ising model,
which has two-phase
behaviour below a critical temperature $\theta_c$. The coexistence curve is
shown in figure~2. In the two-phase region there are two values of the
density which are in equilibrium for each temperature $\theta$.
The same is true for the binary system after carrying out the
transformations mapping it to the reference system. For each
temperature below the critical temperature $T_c$ in the
binary system, the binodal curve corresponds to the coexistence
curve in the reference system. By varying the temperature the binodals will
then form the coexistence surface.

\centerline{[Insert figure~2 about here]}

Before investigating the nature of the coexistence surface it is important
to review some of the properties of the one-component lattice gas in the
two-phase
region. Nearest neighbour lattice gases of dimensionality greater than one
have a critical
density $\rho_c=1/2$, with the critical isochore corresponding to this density.
The properties along the critical isochore are known exactly only in the
two-dimensional case. However, the behaviour in three dimensional systems is
known with sufficient accuracy for our purposes. In this paper we use the
results
obtained via SCOZA.

The equation of the coexistence curve can be written as $\theta=\tau(\rho)$,
for which there are two values of $\rho$ for each $\theta$ below the critical
temperature $\theta_c$. The activity along the critical isochore is
\begin{eqnarray}
\lambda = \exp\left(-{Z\epsilon\over 2k\theta}\right).
\end{eqnarray}
As $\theta\rightarrow\theta_c$ in the two-phase region, the usual power law
expressions are assumed to hold. The coexistence curve in the neighbourhood
of the critical density can be written as
\begin{eqnarray}
\theta_c - \tau(\rho) \sim |\rho - \rho_c|^{1\over \beta},
\end{eqnarray}
where the exponent $1/\beta$ is very close to 3  in three dimensions, so the
coexistence curve is nearly cubic. In SCOZA, $1/\beta=20/7$ \cite{scoza3}.
The heat capacity (constant volume)
diverges as the critical point is approached asymptotically along the
critical isochore, following
\begin{eqnarray}
\Gamma \sim (\theta_c - \theta)^{-\alpha'},
\end{eqnarray}
where $\alpha'>0$ and nearly equal to zero in three dimensions. In SCOZA,
$\alpha'=1/10$ \cite{scoza3}.
Current best estimates for $\beta$ and $\alpha'$ are
$0.3257\pm 0.0025$ and $0.1130\pm 0.0075$ \cite{zinn},
respectively.

\vskip 12pt
\centerline{5.2. Behaviour of a general binary model}
\vskip 12pt

The complete three-dimensional phase diagram for a binary system at constant
pressure is shown in figure~3.

\centerline{[Insert figure~3 about here]}

The coexistence surface separates
the one- and two-phase regions of the system. In a constant temperature plane
the variables are the number densities
of the two components. The two-phase region is bounded by a binodal curve APB,
containing tie lines between points representing the compositions of the two
coexisting phases. The tie lines vanish at point P, called the plait point
(isothermal critical mixing point), where the coexistent phases lose their
separate identities to merge into a single homogeneous phase. In the full
three-dimensional diagram the plait points form a curve on the coexistence
surface. This curve ends up reaching a maximum at the top C, above which
 only one phase is found. C is then a critical
point of the system. We will see that in some cases the binodals are closed
loops involving two plait points.

\vskip 12pt
\centerline{5.3. Widom's model}
\vskip 12pt

The number densities for Widom's model, which is defined by $\epsilon=0$ and
$\phi=+\infty$, are
\begin{eqnarray}
\rho_1 &=& \rho_R \nonumber \\
\rho_2 &=& {1\over 3} \left({q\over 2}-q\rho_R +{\Phi_R\over \epsilon_R}\right)
[\exp(-\beta_R\epsilon_R)-1].
\label{den-widom}
\end{eqnarray}

In order to obtain the binodal curve, shown in figure~4, we
use the reference system coexistence curve values for the energy density,
temperature and number density in equation~\ref{den-widom}.

\centerline{[Insert figure~4 about here]}

The binodal curve {\sl APB} and the line {\sl AB} for $T=0$ separates the
two-phase
region from the remainder of the composition triangle corresponding to the
homogeneous states of the system.
In order to determine the tie line in the binary system we identify the
double-valued density in the reference system for a given temperature. The
corresponding densities in the binodal are coexistent.
The tie line {\sl CD} connects pairs of points on the binodal curve which
represent the compositions of coexistent phases. The longest tie line connects
the coexistent densities ($\rho_1=0,\rho_2=1$) and  ($\rho_1=1,\rho_2=0$),
corresponding to $\theta=0$ in the reference system.
The tie lines vanish at the point {\sl P}, where the conjugate phases merge
into a single homogeneous phase. Thus {\sl P} is a critical point achieved by
changing composition at fixed temperature. {\sl P} is frequently called a
plait point in order to distinguish it from the usual critical point achieved
by changing temperature. The coordinates $\rho_1^*,\rho_2^*$ of the plait
point are
\begin{eqnarray}
\rho_1^* = {1\over 2}, ~~~ \rho_2^* = {1\over 3}\left({\Phi_c\over \epsilon_R}\right)
[\exp(-\beta_c\epsilon_R)-1]=0.1968242
\label{plait-widom}
\end{eqnarray}

The binodal curve  is an isothermal section of
a more general three-dimensional diagram. For this model the number
density and all the other quantities derived are independent of the
temperature $T$. This happens because the molecular interaction potential
is either $0$ or
$+\infty$, so the total potential energy of this system never deviates from
$0$, i.e. there is no energy scale. Nevertheless, the model exhibits phase
coexistence and an associated plait point.

\vskip 12pt
\centerline{5.4. Clark's model}
\vskip 12pt

For this model, $\epsilon = 0$ and $\phi$ is finite. The number densities
$\rho_1$ and $\rho_2$ are
\begin{eqnarray}
\rho_1 &=& \rho_R \nonumber\\
\rho_2 &=& {1\over 3}\left[{q\over 2}A + N{q\over 2}(C-A) + N_{11}(A + B - 2C)\right]
\label{den-clark}
\end{eqnarray}
where $A, B$ and $C$ were defined by equation~\ref{abc} when deriving the grand
partition function.

Equations~\ref{mapping} and \ref{den-clark} are the mapping between Clark's
model and the
lattice gas reference system. In order to obtain the binodal curve, we
determine
the activities and the number densities for the binary model by using the
energy density, number density and temperature values of the reference system
coexistence curve.

\centerline{[Insert figure~5 about here]}

Figure~5 shows the isothermal binodal curves in the $(\rho_1,\rho_2)$
plane, corresponding to $\phi=0.5$ and temperatures $T=0.1;0.2;0.247162$.
The coexistence surface, figure~3, is the set of all such
binodal curves, which close to form a dome whose apex marks the critical
point ($T_c=0.247173$). Each isotherm in the $(\rho_1,\rho_2)$ plane, has an
upper and a lower plait point. These plait points form a plait point curve
that ends at the critical point $T_c$.

\vskip 12pt
\centerline{5.5. Neece's model}
\vskip 12pt

For this model $\epsilon$ is nonzero and $\phi=+\infty$. The number densities
are
\begin{eqnarray}
\rho_1 &=& \rho_R \nonumber\\
\rho_2 &=& {1\over 3}\left({q\over 2} + Nq + N_{11}\right)\left[1-
\exp\left(-{\epsilon_R\over k\theta}-{\epsilon\over kT}\right)\right].
\label{den-neece}
\end{eqnarray}
If $\epsilon<0$ there is an attraction between nearest neighbours on the primary
lattice. In our model $|\epsilon_R|=|\epsilon|=1$. The second term in brackets
 in the expression for $\rho_2$ is never
 zero for $T>\theta_c$. If $T=\theta_c$, then this factor is zero and
 $\rho_2=0$ at $\rho_1=1/2$. If $T<\theta_c$, the second factor is zero for some
 value of $\theta$ which in turn corresponds to two values of $\rho_1$.
 The binodal curve for $\epsilon <0$ is shown in figure~6 at
 different temperatures. The critical temperature is $T_c=1.1299$, represented
 by curve {\it C}. From the inset of figure~6, which is a zoom of curve D, we
 can see that there are two separate regions
 corresponding to the homogeneous states of the system for $T<\theta_c$. Also,
 in this case there is no plait point since there is no tie line of vanishing
 length.

 \centerline{[Insert figure~6 about here]}

 For $\epsilon>0$, the binodals as a function of temperature are shown in
 figure~7. The plait points lie along $\rho_1=1/2$.

 \centerline{[Insert figure~7 about here]}

\vskip 12pt
\centerline{5.6. Wheeler's model}
\vskip 12pt

 For this model, $\epsilon < 0$ and $\phi$ is finite (attractive or repulsive).
 The number densities $\rho_1$ and $\rho_2$ are given by

 \begin{eqnarray}
 \rho_1 &=& \rho_R  \nonumber \\
 \rho_2 &=& ({q\over 2})A + q(C-A)\rho_R + (A+B-2C){\Phi_{11}\over \epsilon_R}
 \end{eqnarray}
 where $A, B$ and $C$ are defined by equation~\ref{abc}.

 \centerline{[Insert figure~8 about here]}

 Figure~8 shows the binodal curves for $\Phi<0$ and $\Phi>0$ at different
 temperatures. Curve A is for $T=1.11$, curve B is for $T=1.25$ and curve
 C for $T=1.336$ for attractive and repulsive interactions. The slope of
 the tie lines is positive for
 $\Phi<0$ and negative for $\Phi>0$. At $T=T_c=1.336498$ the binodal
 shrinks to a point. In a 3-D diagram, including the temperature, it
 corresponds to a dome that separates the two-phase region from the
 homogeneous one.

\vskip 12pt
 \centerline{5.7. Wheeler's oriented model}
 \vskip 12pt

 In this model particles of the same species (1-1 or 2-2) do not interact.
 The 1-2
 interaction depends upon the orientation of the particle in the secondary
 cell. If the 1-2 bond involves the special contact point of this particle
 then the energy is $U_2<0$ (attractive), otherwise it is $U_1>0$ (repulsive).

 \centerline{[Insert figure~9 about here]}

 Figure~9 shows the coexistence curve for $r=|U_2|/U_1=0.189$ (curve A),
 $0.18$ (curve B), and $0.17$ (curve C). Each
 curve forms a closed loop with a lower and upper critical temperature. The
 ratios $T_u/T_l$ of upper to lower critical solution temperature are
 respectively 1.161, 1.467 and 1.721 for these curves. Our results are in
 close agreement with those of Wheeler \cite{scesney}, using low temperature
 series expansions and with the experimental data.

 In figure~10 we show the experimental coexistence curves for solutions of
 glycerol and guaiacol (circles) \cite{mcewan} and m-toluidine (squares)
 \cite{parvatiker}. These curves are qualitatively similar to the closed
 loops obtained from Wheeler's oriented model.

 \centerline{[Insert figure~10 about here]}

\vskip 12pt
 \centerline{\bf 6. Concluding remarks}
 \vskip 12pt

 We study binary systems through various versions of the decorated lattice
 model using the self-consistent Ornstein-Zernike approximation.
 We obtain the binodals, plait points, and the critical temperatures for
 the models introduced by Widom, Neece, Clark and Wheeler.
 SCOZA provides a very useful way of getting accurate results for binary
 systems that can be mapped to the lattice-gas Ising model.

 We also study Wheeler's oriented model, obtaining its coexistence curve for
 various interaction energies. The results are consistent with experimental
 data and low temperature series expansions.

 In order to analyse the algebraic degree of the binodal at the plait point
 we can use the idea proposed by Widom \cite{widom67}.
 It consists in measuring the vertical
 distance $\Delta \rho_2$ from the binodal curve to a line drawn tangent to the
 binodal at the plait point. By combining the number densities with the energy
 density and the coordinates of the plait point, we find that
 \begin{eqnarray}
 \Delta\rho_2 \sim |\rho_1 - {1\over 2}|^{(1-\alpha')\over \beta}.
 \end{eqnarray}
 The modification of the critical exponent by the factor $(1-\alpha')$ is an
 example of `renormalization' discussed by Fisher \cite{fisher}.
 The SCOZA values for $\beta$ and $\alpha'$ are 7/20 and 1/10,
 respectively, providing 18/7 for the algebraic degree of the binodal.

 These results can be extended to study ternary systems as well as pure fluids.
 It is not difficult to derive the energy, heat capacity, pressure for all the
 models studied in this paper.

 A.G.D. is grateful to CAPES and Fapemig (Brazil) for financial support.
 A.G.D. is also grateful to Ann and the staff of SUNY at Stony Brook for
 their hospitality and kindness during her stay.
 G.S. gratefully acknowledges the support of the 
 Division of Chemical Sciences, Office of Basic Energy Sciences, Office of
 Energy Research, US Department of Energy. The authors thank F. Raineri for
 his assistance.

 \newpage

 \begin{table}
 \caption{Interactions between molecules in adjacent cells for the different
 decorated models.}
 \begin{center}
 \begin{tabular}{|c|c|c|}\hline
 Model    & Energy between type 1 molecules & Energy between unlike
 molecules \\ \hline
 Widom    & $\epsilon=0$                 & $\phi=+\infty$       \\ \hline
 Clark    & $\epsilon=0$                 & $\phi>0$ or $\phi<0$ \\ \hline
 Neece    & $\epsilon>0$ or $\epsilon<0$ & $\phi=+\infty$       \\ \hline
 Wheeler  & $\epsilon>0$                 & $\phi>0$ or $\phi<0$ \\ \hline
 Wheeler oriented & $\epsilon=0$         & $U_1>0$ or $U_2<0$   \\ \hline
 \end{tabular}
 \label{table1}
 \end{center}
 \end{table}

 \newpage

 \noindent{\bf Figure Captions}
 \vspace{1em}

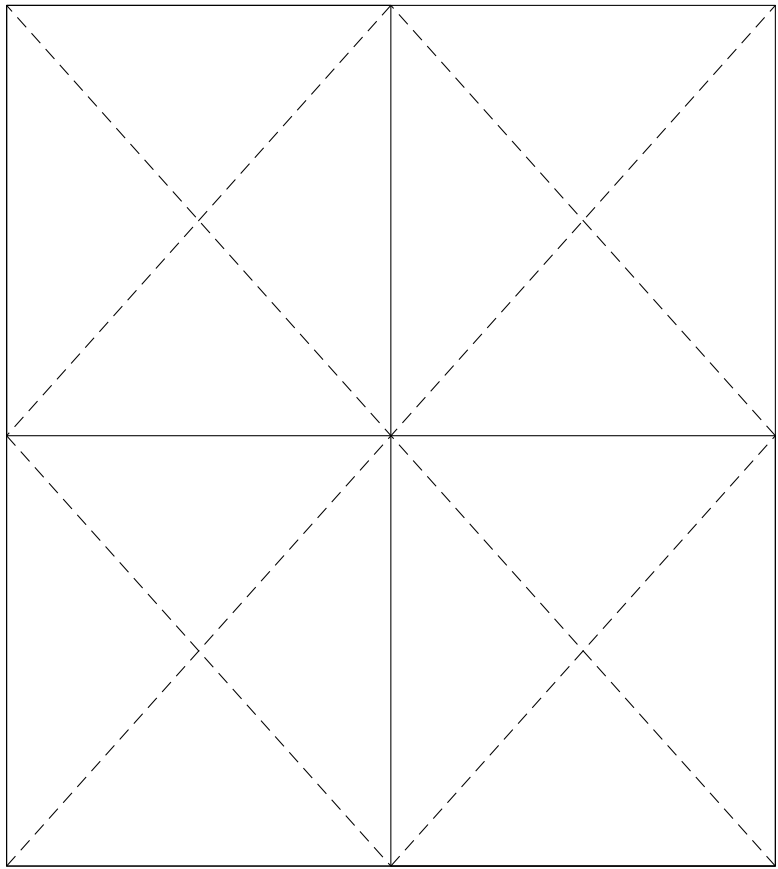
 \noindent Figure~1. Example of a decorated lattice in two dimensions.
 The solid squares represent the primary cells while the dashed diamonds
 represent the secondary cells.
 \vspace{1em}

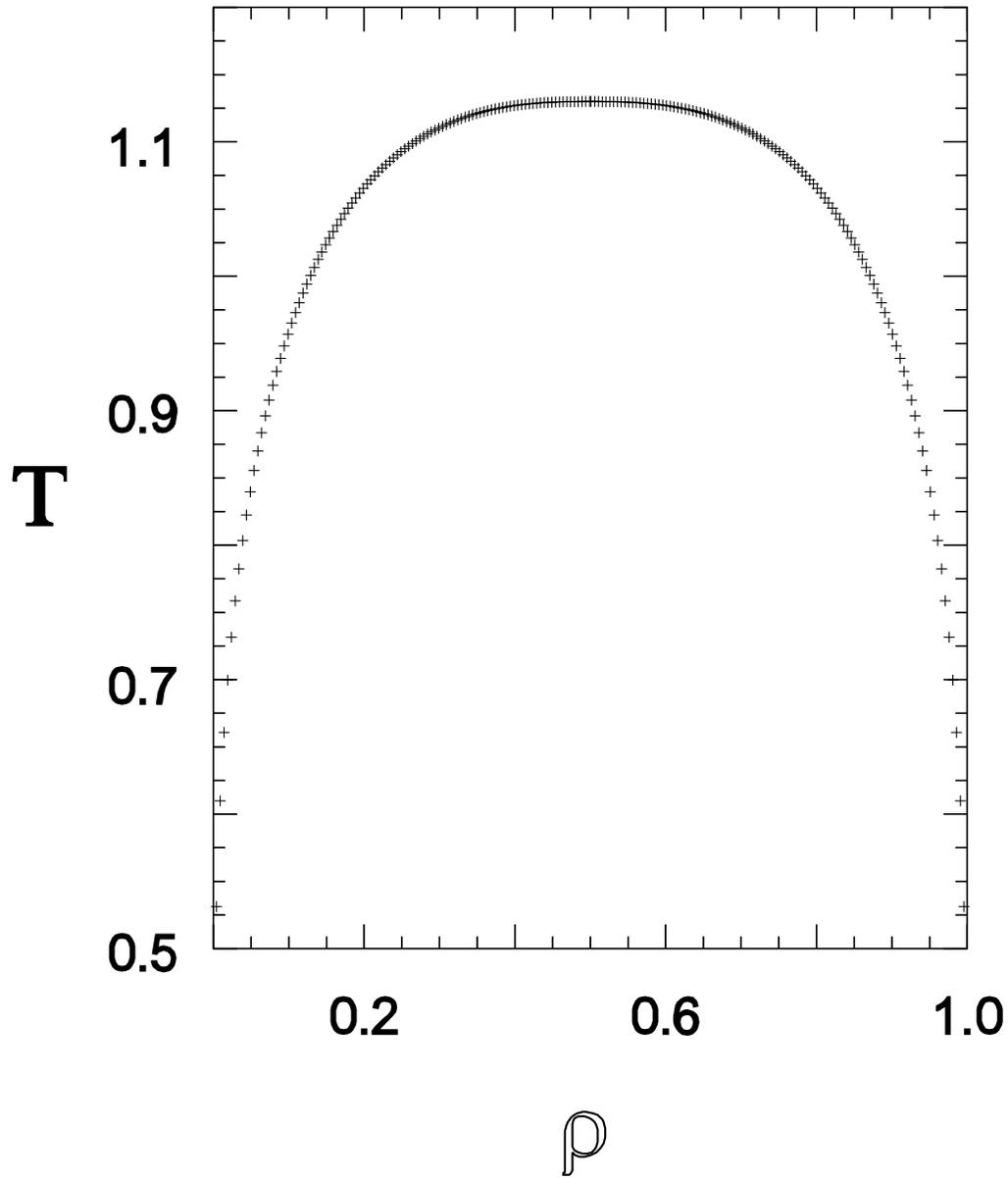
 \noindent Figure~2. Coexistence curve for the lattice gas Ising model using
 SCOZA results.
 \vspace{1em}

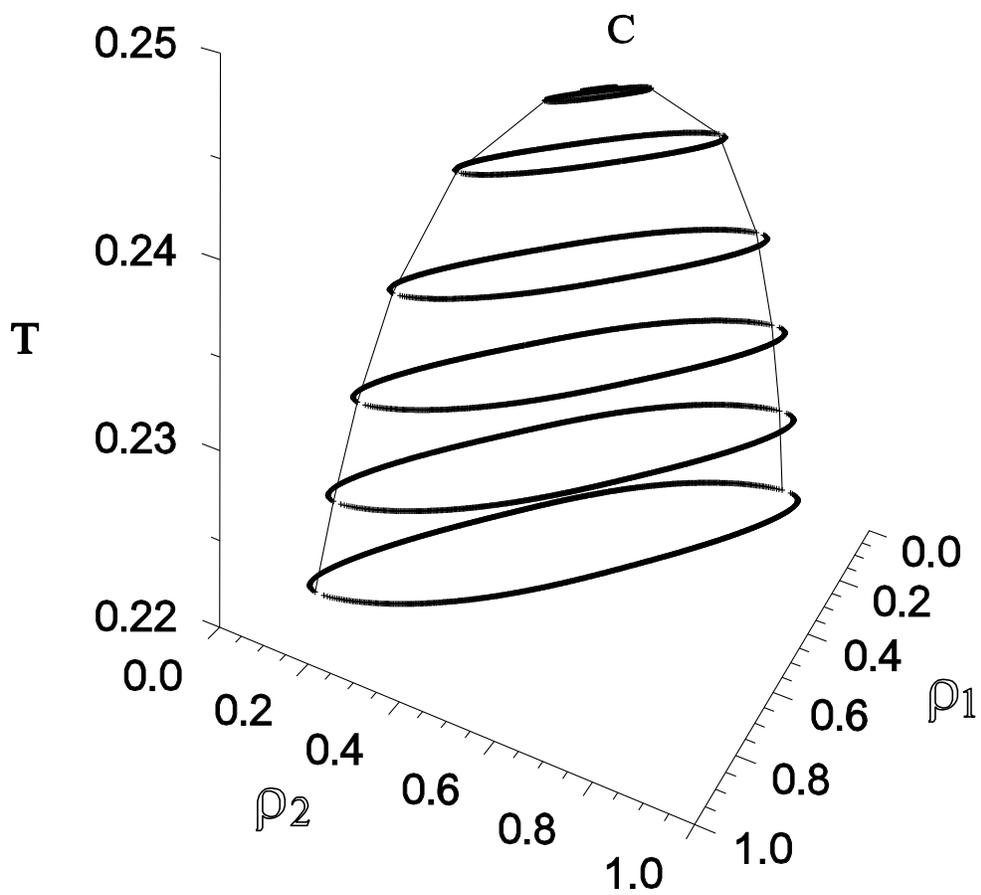
 \noindent Figure~3. Coexistence surface for a binary mixture using the results
 for Clark's model. The top C marks a critical point.
 \vspace{1em}

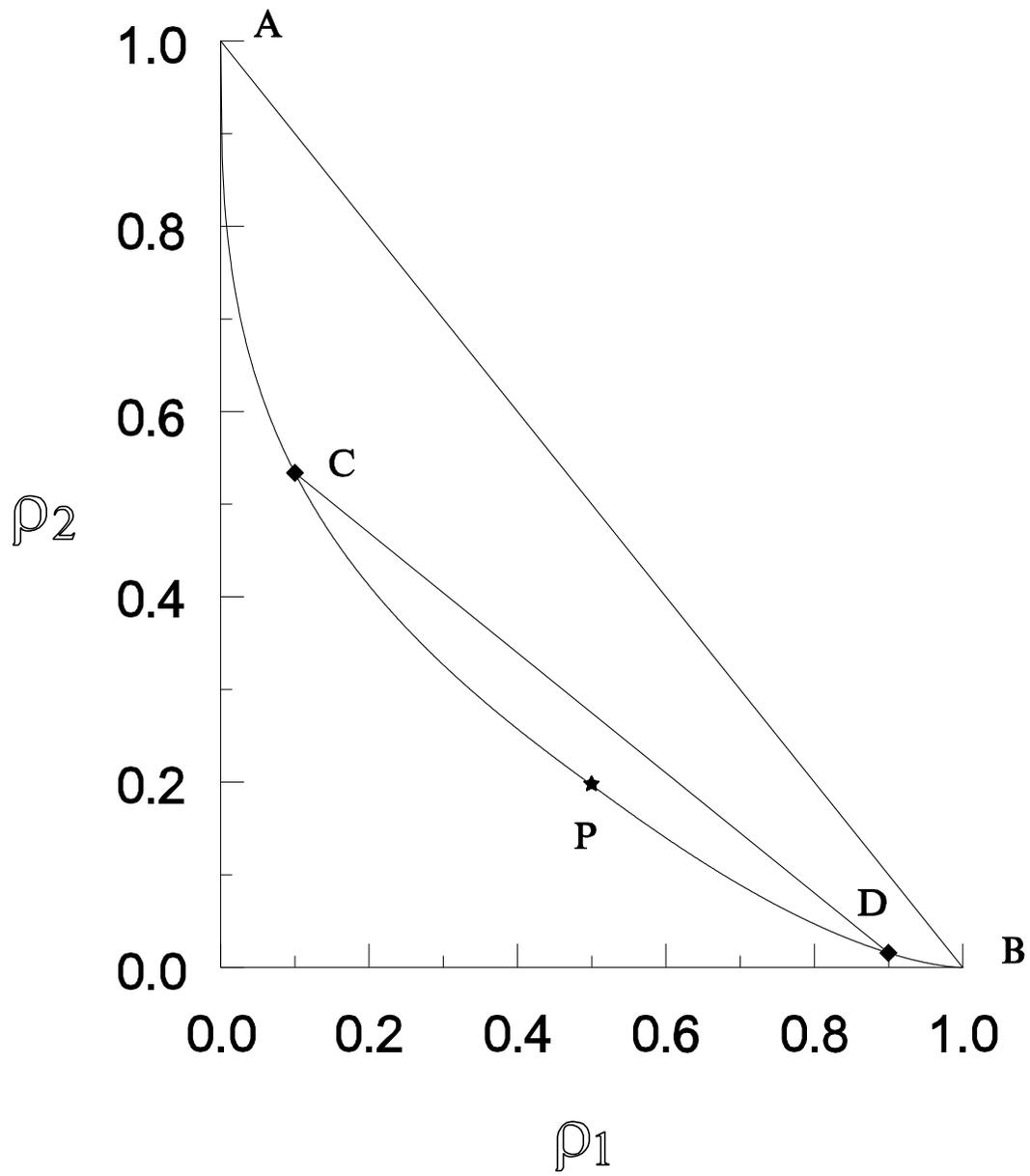
 \noindent Figure~4. Binodal curve for Widom's model. {\sl CD} is a tie line
 and {\sl P} is the plait point. The composition variables $\rho_1$ and
 $\rho_2$ are the number densities of the two components.
 \vspace{1em}

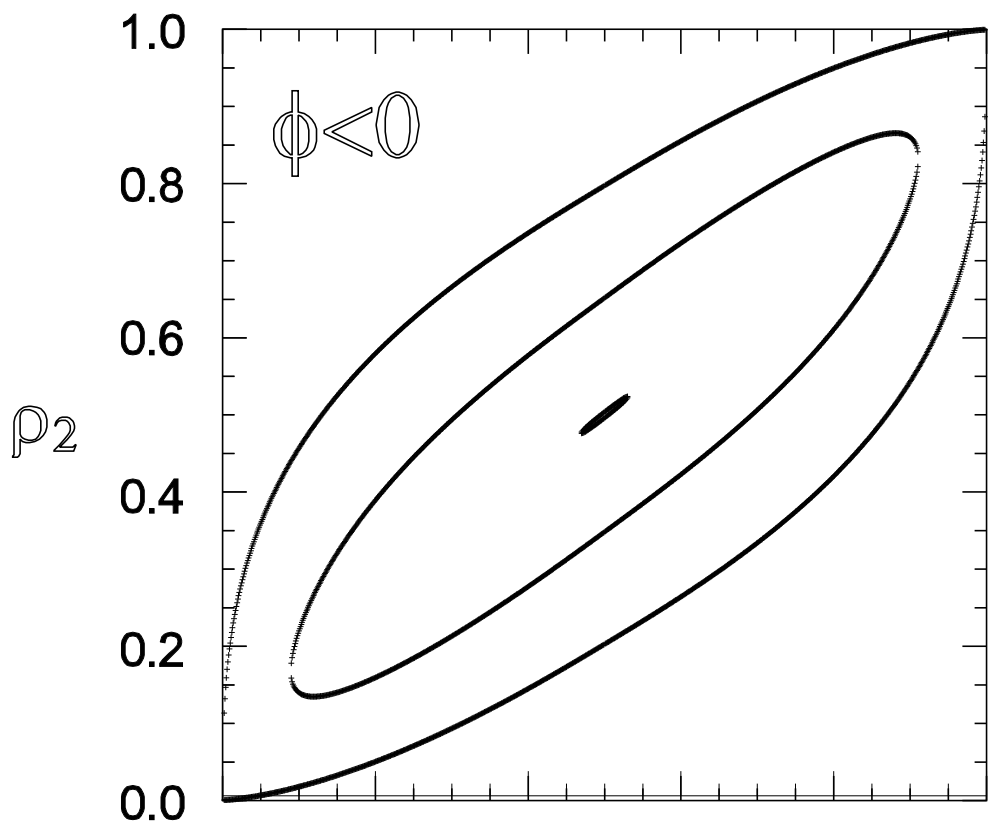
 \noindent Figure~5. Binodal curve for Clark's model for attractive ($\Phi>0$)
  and repulsive ($\Phi<0$) interactions. The outer binodal corresponds to
   $T=0.1$; the intermediate one corresponds to $T=0.2$ and the inner curve
    is for $T=0.247$, very close to $T_c=0.247173$.
    \vspace{1em}

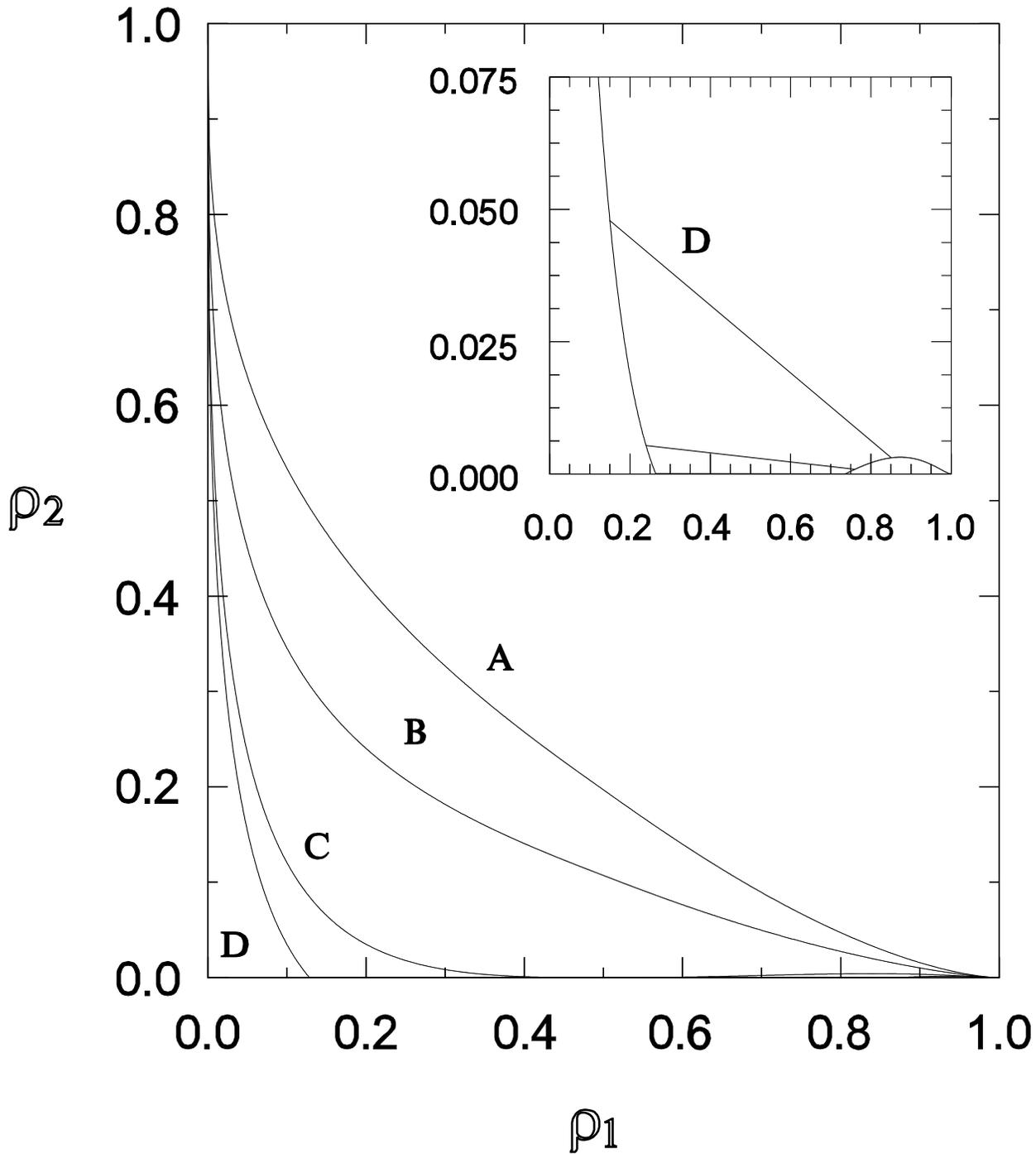
    \noindent Figure~6. Binodal curve for Neece's model with $\epsilon<0$. Curve
    {\it A} corresponds to $T\rightarrow\infty$; curve {\it B} is for $T=2.0>T_c$;
    the critical temperature $T_c=1.12989$ is represented by curve {\it C}; and
    $T=1.0<T_c$ at curve {\it D}. The inset is a zoom of curve {\it D}.
    \vspace{1em}

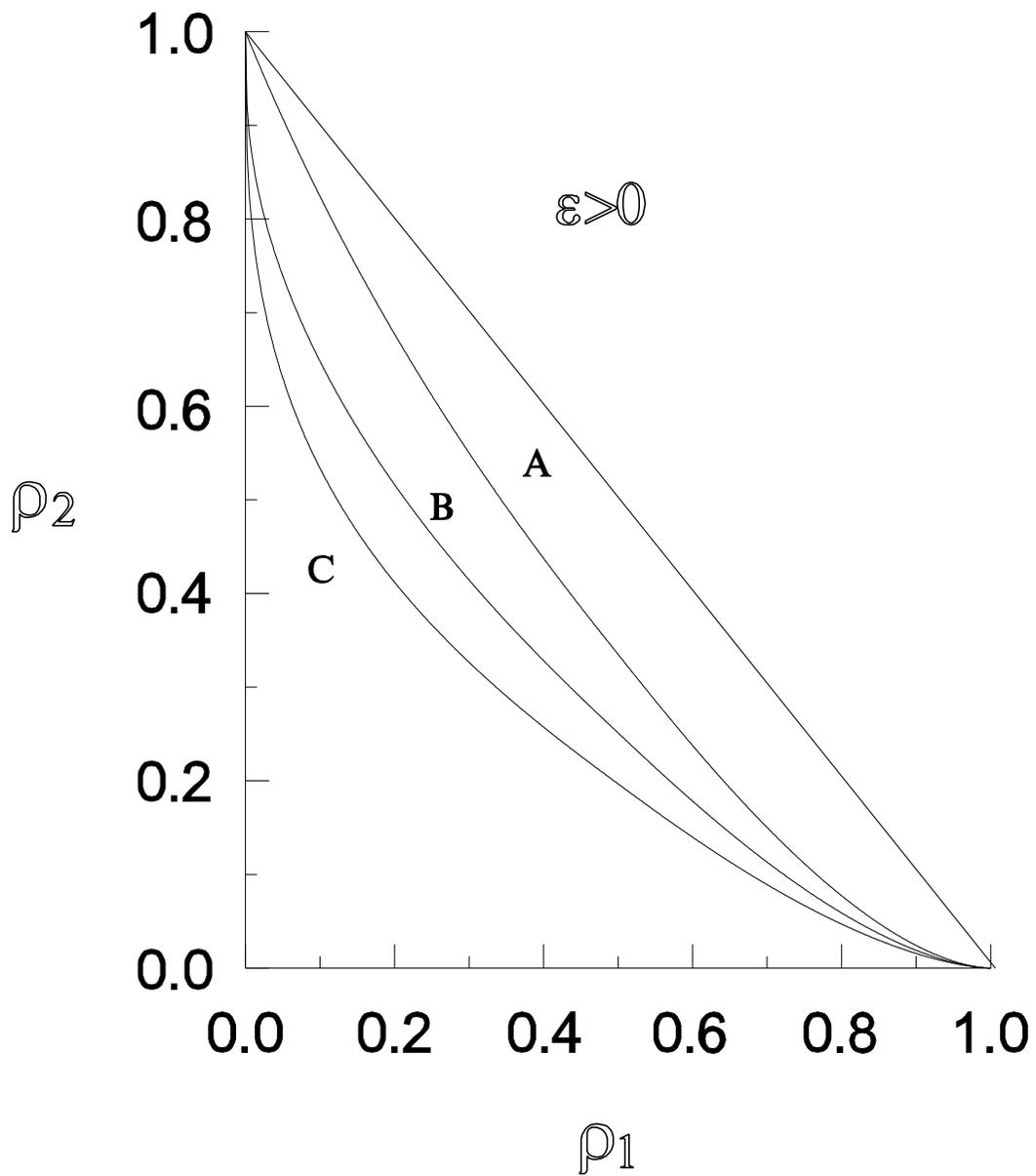
    \noindent Figure~7. Binodal curve for Neece's model with $\epsilon>0$. Curve
    {\it A} corresponds to $T=0$; curve {\it B} is for $T>0$; $T\rightarrow\infty$
    is represented by curve {\it C}.
    \vspace{1em}

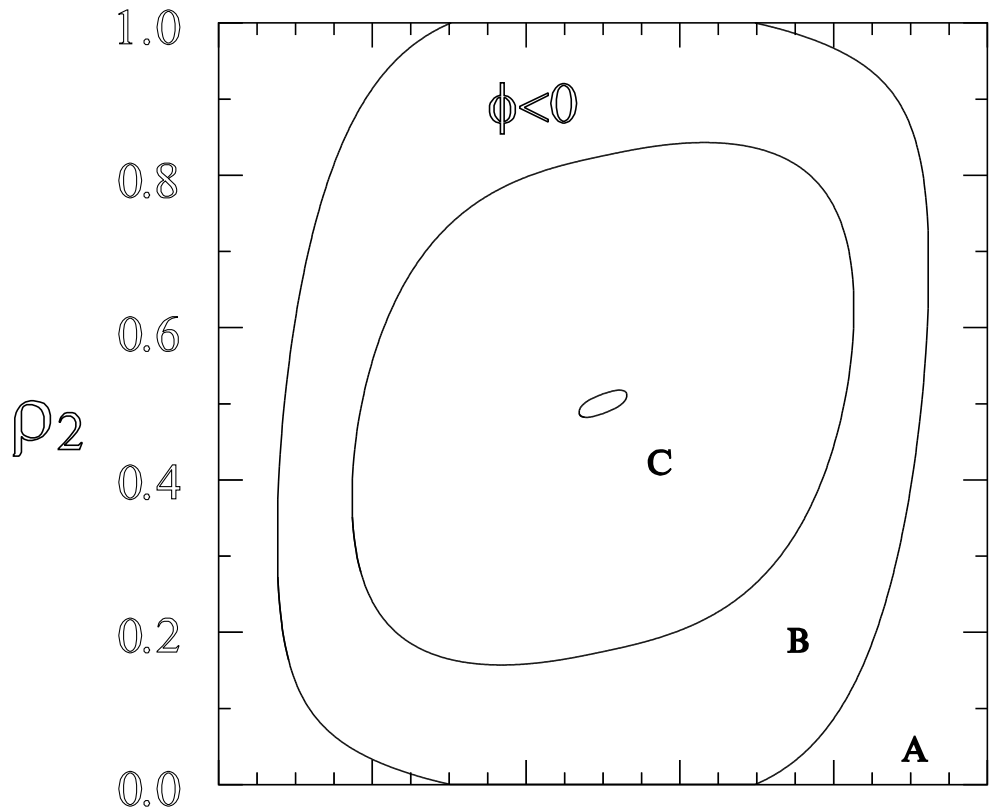
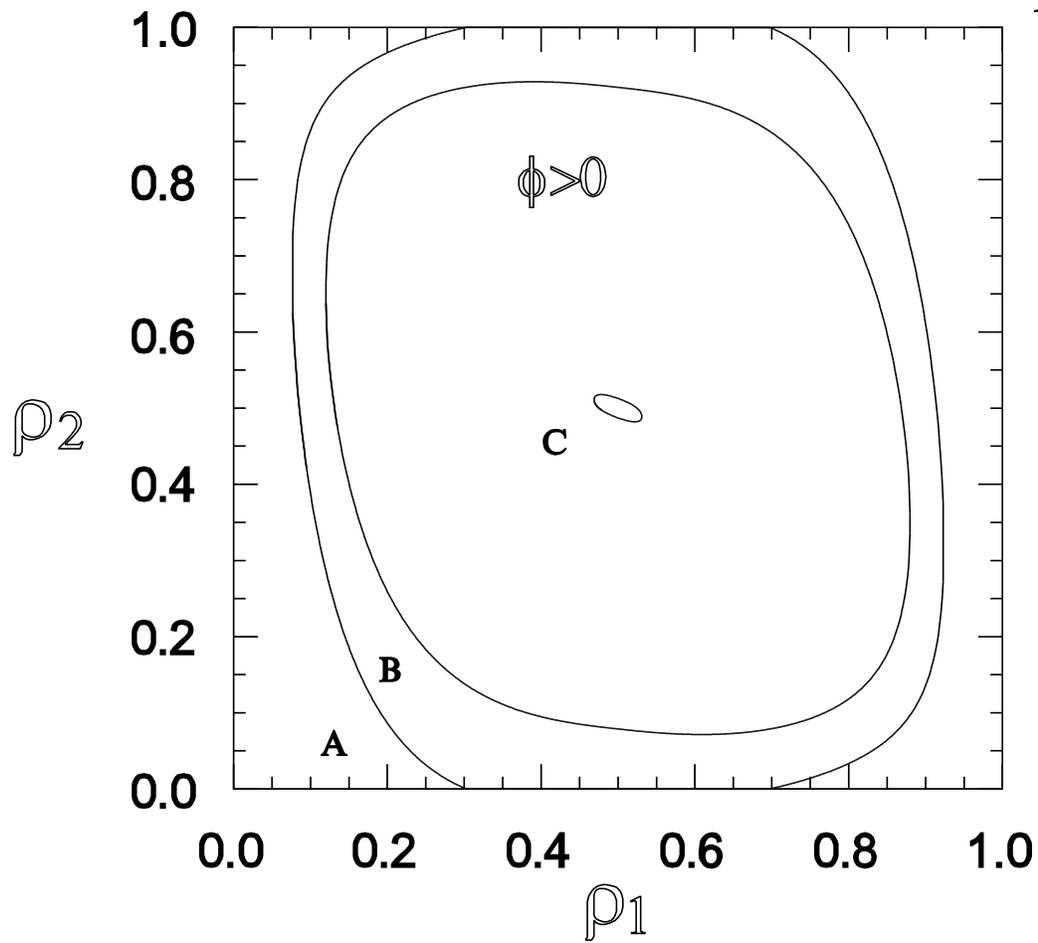
    \noindent Figure~8. Binodal curve for Wheeler's model for
    repulsive ($\Phi<0$) and attractive ($\Phi>0$) interactions.
     Curve A is for $T=1.11$,
    curve B is for $T=1.25$ and curve C for $T=1.336$ for both diagrams.
    The critical temperature is $T=T_c=1.336498$.

    \vspace{1em}

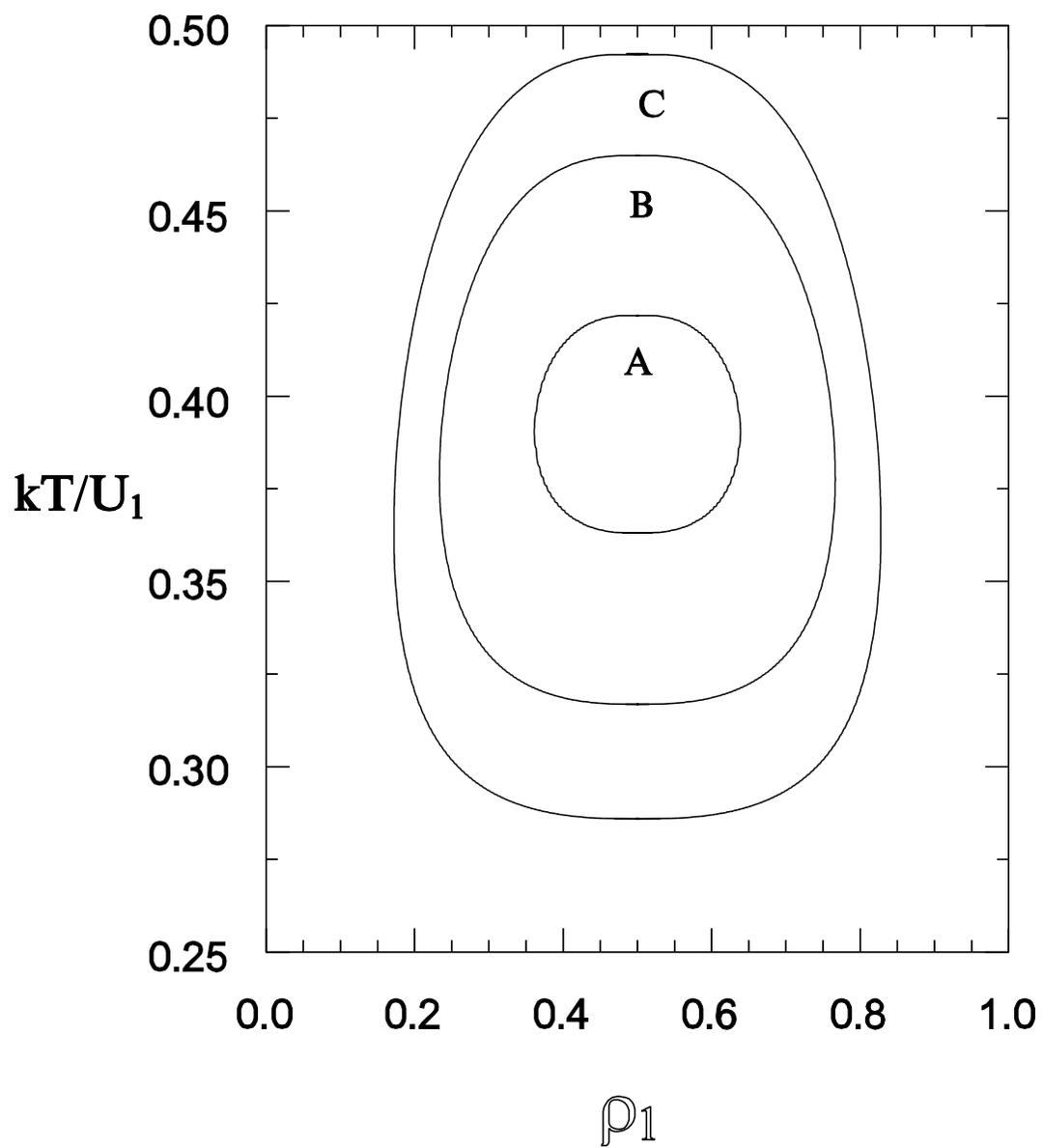
    \noindent Figure~9. Coexistence curves for Wheeler's oriented model in the
    temperature-mole fraction diagram. Curves {\it A}, {\it B} and {\it C}
    correspond to $r=0.189, 0.18, 0.17$, respectively. 

    \vspace{1em}

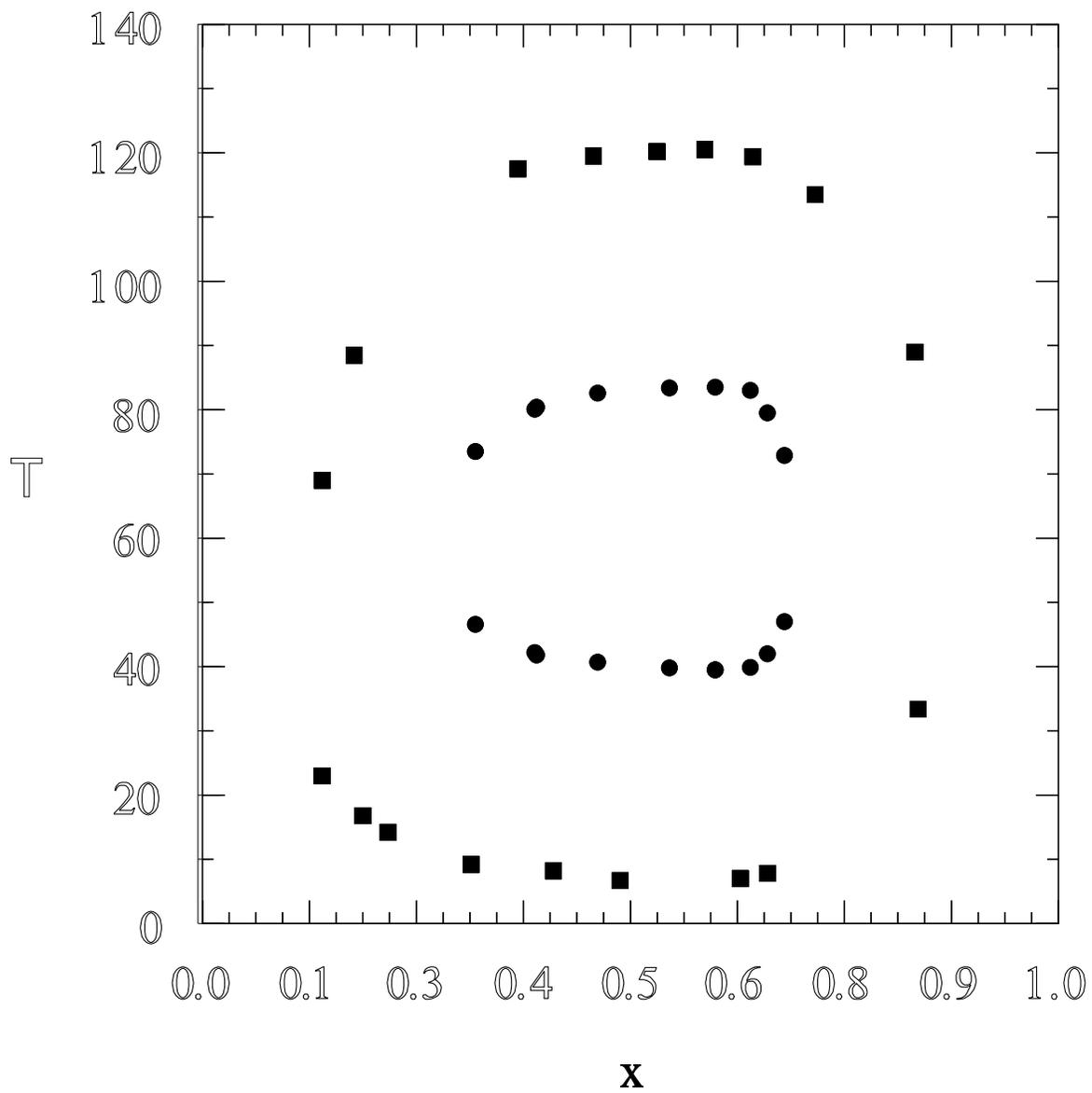
    \noindent Figure~10. Temperature-mass fraction coexistence curves for solutions
    of glycerol with guaiacol (circles) and m-toluidine (squares).
    Temperature is in Celsius.

    
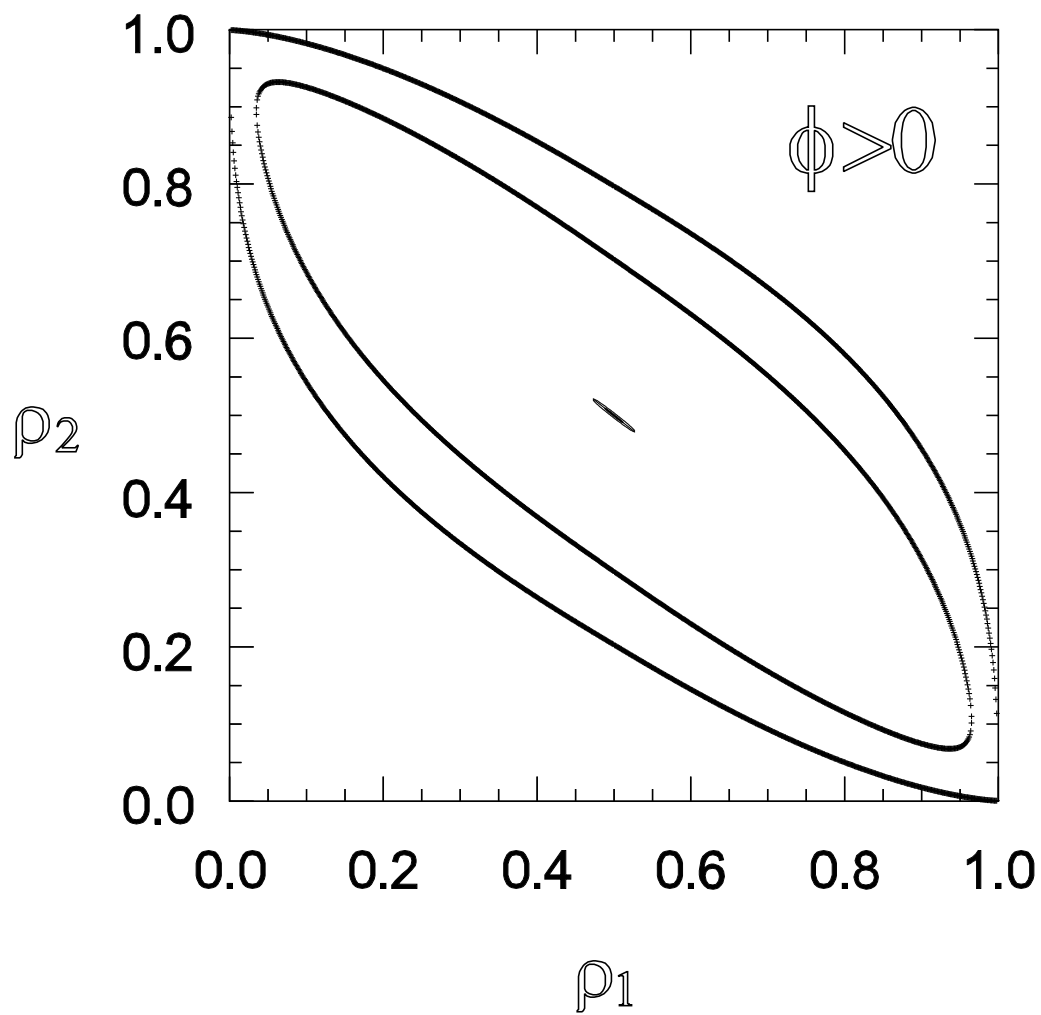
\end{document}